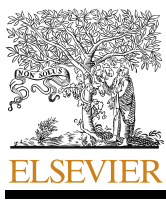
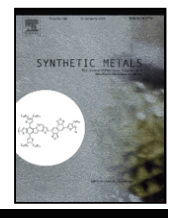

# Optimizing transparent electrodes: Interplay of high purity SWCNTs network and a polymer

Sara Joksović [a], [*], Jovana Stanojev [a], Nataša Samardžić [b], Branimir Bajac [a]

[a] *BioSense Institute, University of Novi Sad, Dr Zorana Đinđića 1, Novi Sad 21000, Serbia*
[b] *Faculty of Technical Sciences, University of Novi Sad, Novi Sad 21000, Serbia*

ABSTRACT

The discovery of transparent electrodes led to the development of optoelectronic devices such as touchscreens, infrared (IR) sensors, etc. Carbon nanotubes (CNTs) have been a potential replacement for ITO due to their exceptional properties, especially in the IR region. In this work, we present the development of a CNT-polymer composite thin film that exhibits outstanding transparency across visible and IR spectra prepared by layer-by-layer (LbL) technique. This approach ensures uniform integration and crosslinking of CNTs into lightweight matrices, and also represents a cost-effective method for producing transparent electrodes with remarkable properties. The produced films achieved a transparency above 80 % in the UV/VIS range and approximately 70 % in the mid-IR range. The sheet resistance of the fabricated thin films was measured at about 4 kΩ/sq, showing a tendency to decrease with the number of bilayers. In this work we have investigated electrical properties and transport mechanisms in more detail with computational analysis. Computational analysis was performed to better understand the electrical behavior of nanotube-polymer junctions in the interbundle structure. Based on all results, we propose that the transparent electrodes with 4 and 6 bilayers are the most optimal structures in terms of optical and electrical properties.

## 1. Introduction

In the domain of optoelectronic devices, transparent conducting electrodes (TCEs) play a crucial role, requiring exceptional optical transparency and electrical conductivity [1–3]. Traditional materials like indium tin oxide (ITO) offer superior optoelectronic properties but come with drawbacks such as high indium cost, expensive deposition methods [4] and brittleness, making them less suitable for the next generation of optoelectronic devices (e.g. flexible optoelectronics) [2,3]. Additionally, the fact that ITO has limited transparency in certain spectra (e.g. near UV and mid-IR) makes it difficult to use for infrared devices [5]. For this reason, there is a need for other materials with high transmittance in a wide range of electromagnetic spectra, with excellent electrical properties for application as a conductive thin layer.

Carbon nanotubes (CNTs) discovered by S. Ijima [6] in 1991, are a promising alternative to ITO [7–9]. Single-walled carbon nanotubes (SWCNTs) stand out for their extraordinary mechanical [10], electrical [10,11] and thermal [10,11] properties and therefore, have been extensively studied for various applications, such as optoelectronic devices [1] and different types of biosensors [12]. As a material for transparent conductive electrodes, SWCNTs offer excellent optical properties and flexibility, crucial for the further development of flexible electronics [5,13]. SWCNTs have very good transparency in UV/VIS spectra, but especially in the mid-IR range, and are therefore a good candidate material for transparent electrodes in the mid-IR range.

The great potential of transparent conductive electrodes based on SWCNT was first introduced in 2004 by Saran et al. [14] and Wu et al. [15]. Up to now, there are numerous papers about the SWCNT applied as a transparent electrode [16]. SWCNT film with the sheet resistance of 50 Ω/sq and 100 Ω/sq with the transparency of 60 % and 70 % respectively was reported by Contreras and coworkers [17]. They investigated the transparency of prepared SWCNT film in the infrared (IR) range and found that such films showed a high transparency and possible application in photovoltaic (PV) devices. Yu et al. [18] treated SWCNT in order to increase transparency and conductivity and developed thin SWCNT films with the transmittance of 88 % and 6 kΩ/sq sheet resistance. In another work, Kuan-Ru Chen et al. [19] investigated different methods for depositing SWCNT on polymer substrates. They developed films with a sheet resistance of about 1 kΩ/sq and transmittance above 85 %. Also, David S Hecht [20] reported excellent results of CNT

[*] Corresponding author.
*E-mail address:* sara.joksovic@biosense.rs (S. Joksović).

films with the high transparency of 90.9 % and low sheet resistance of 60 Ω/sq.

From the literature, the SWCNT based transparent electrodes have been intensively studied for a wide range of applications, but it is important to point out that they have been investigated mostly in the UVvisible range. In this work, we present an optimized transparent electrode for the application in the mid-IR range with the emphasis on the deeper understanding of electrical behavior of the SWCNT network.

In light of the aforementioned, the analysis of electrical properties delves to in-depth understanding of developed material electrical properties, conductivity mechanisms of SWCNT network in deposited film form, and the influence of an added polymer assisting the deposition. The conductivity of a single SWCNT is directly linked to the length of the conductive path, but on the other hand nanotube network contains both intertube and interbundle junctions, with the polymer layers introducing additional junctions with the SWCNT. Transport properties of such film are consequently influenced by various mechanisms, including scattering, thermally assisted tunneling and strongly dependent on the energy profile of junctions. This study was necessary to understand how to tailor the transparent thin film electrode, with the primary aim to balance the tradeoff between electrical and optical properties.

## 2. Experimental study

In this study, we present the fabrication and characterization of a mid-IR transparent electrode based on high-purity SWCNTs. The thin films were prepared on a glass substrate using the layer-by-layer (LbL) technique. The structural properties of a prepared films were investigated with the Horiba XploRA Plus Raman microscope (λ = 532 nm, 9 mW power, 5 exposures in 10 s, ×100 magnification) and Thermofisher Scientific Apreo 2 C low-vac High-Resolution Scanning Electron Microscope (×50 000 and ×100 000 magnification). Optical properties were examined by UV/VIS Jasco V-750 (visible range: 400–900 nm, scan rate 400 nm/min) and Jasco FT/IR-6600 (mid-IR range: 2500 – 3500 $cm^{-1}$). Electrical characterization was investigated in detail with Hewlett Packard 3457 A sheet resistance multimeter, Hall effect measurement system Ecopia HMS-3000, Palmsens 4, and Keithley 4200A-SCS parameter analyzer (range: from −1 V to 1 V, step by 0.01 V). The characterization results are presented and discussed in the Results and discussion section.

### 2.1. Experimental procedure

Layer-by-layer (LbL) is a cost-effective self-assembly technique for depositing high-quality, uniform thin films using single-walled carbon nanotubes. The process involves alternating layers of positively charged polyethyleneimine (PEI) and negatively charged carboxyl functionalized SWCNTs (Sigma Aldrich, purity > 90 %). Soda lime glass substrates were prepared following the procedures from our previous publications [21,22], with a minor modification of sonication for 10 min before each deposition to prevent SWCNT agglomeration. PEI-SWCNT thin films with 2, 4, 6, 8, and 10 bilayers were fabricated on glass and exposed to drying (120 °C for 10 min) and thermal treatment (300 °C for 1 h) to investigate the influence of thermal treatment on the structure and properties. Refer to Table 1 for sample labeling details.

## 3. Results and discussion

### 3.1. Structural characterization

#### 3.1.1. Raman spectroscopy

Raman spectroscopy (green laser, λ = 532 nm) provided insight into the molecular structure of the 1D carbon materials. Fig. 1 represents PEI-SWCNT thin films with 10 bilayers before (black line) and after (red line) thermal treatment. The most intensive peak, the G-band, originating from $sp^2$ hybridized carbon atoms, displayed two sub-bands [23,24], $G^-$ and $G^+$. According to the literature [24,25] and the shape of the peaks, semiconducting behavior dominates in this material. Peaks at ~1572 $cm^{-1}$ ($G^-$) and ~1595 $cm^{-1}$ ($G^+$) corresponded to vibrations in circumferential and axial directions, respectively [23,24]. The D band at ~ 1346 $cm^{-1}$ originates from the scattering on the $sp^3$ hybridized carbon atoms, indicating disorder in the structure, such as amorphous carbon atoms, vacancies, heteroatoms etc. The low intensity of the D band suggested high SWCNT quality [23–27]. An additional peak at ~ 2679 $cm^{-1}$ is a G' band and it is a second-order overtone of the D band originating from the defects [24–26]. At the low values of the Raman shift, the observed peak at ~ 175 $cm^{-1}$ originates from the Radial Breathing Mode (RBM), a specific vibrational mode only for SWCNTs [24,25,27]. The Raman spectra show that there is no significant difference between the sample before and after thermal treatment, indicating that the 1D structure of the nanotubes is undisrupted and stable [28].

**Table 1**
Labeling of the samples.

| Component 1 | Component 2 | Number of bilayers | Sample name | |
|---|---|---|---|---|
| | | | Temperature of 120 °C, 10 min | Temperature of 300 °C, 60 min |
| PEI | SWCNT-COOH | 2 | PEI-SWCNT (2) | PEI-SWCNT (2) 300 |
| PEI | SWCNT-COOH | 4 | PEI-SWCNT (4) | PEI-SWCNT (4) 300 |
| PEI | SWCNT-COOH | 6 | PEI-SWCNT (6) | PEI-SWCNT (6) 300 |
| PEI | SWCNT-COOH | 8 | PEI-SWCNT (8) | PEI-SWCNT (8) 300 |
| PEI | SWCNT-COOH | 10 | PEI-SWCNT (10) | PEI-SWCNT (10) 300 |

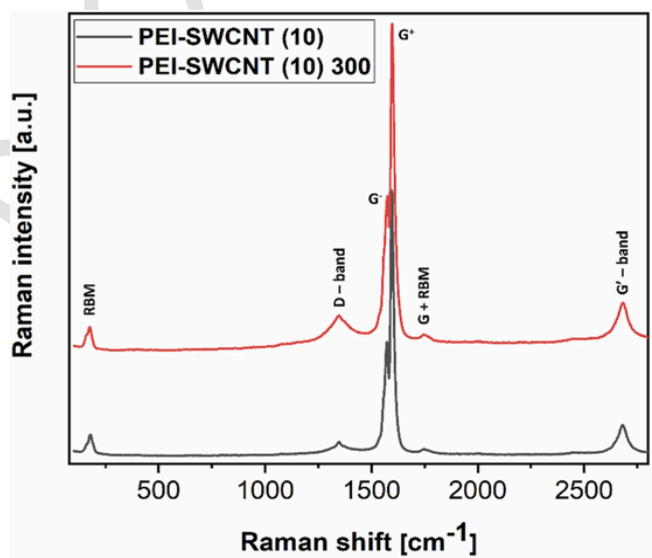

**Fig. 1.** Raman spectra of PEI-SWCNT (10) and PEI-SWCNT (10) 300.

#### 3.1.2. High resolution scanning electron microscopy (HR SEM)

HR SEM microscopy was used to gain information about the morphology and distribution of the SWCNTs in the composite thin films before and after thermal treatment. Fig. 2a and 2b show SEM images of the surface of PEI-SWCNT (10) and PEI-SWCNT (10) 300 films deposited on the soda lime glass substrate. In both cases, the distribution of SWCNTs is relatively uniform along the surface. Before thermal treatment, it is observed that the SWCNTs are well interconnected forming the network along the surface of the sample (Fig. 2a), with notable surface porosity. After 300 °C for 1 h the structure seems denser, with reduced porosity than before (Fig. 2b). It is assumed that by exposing PEI polymer to 300 °C, polymer chains start to flow and the material itself starts to transition into a melted state, reducing its viscosity, which allows present pores in the system to be filled improving the density of the electrode. This temperature was chosen based on TG/DSC analysis, which exactly shows the beginning of polymer state transition without decomposition (Fig. 2c). This process not only improves the density, but increases the average proximity of SWCNTs, and the probability of direct contact which should reflect on electrical properties.

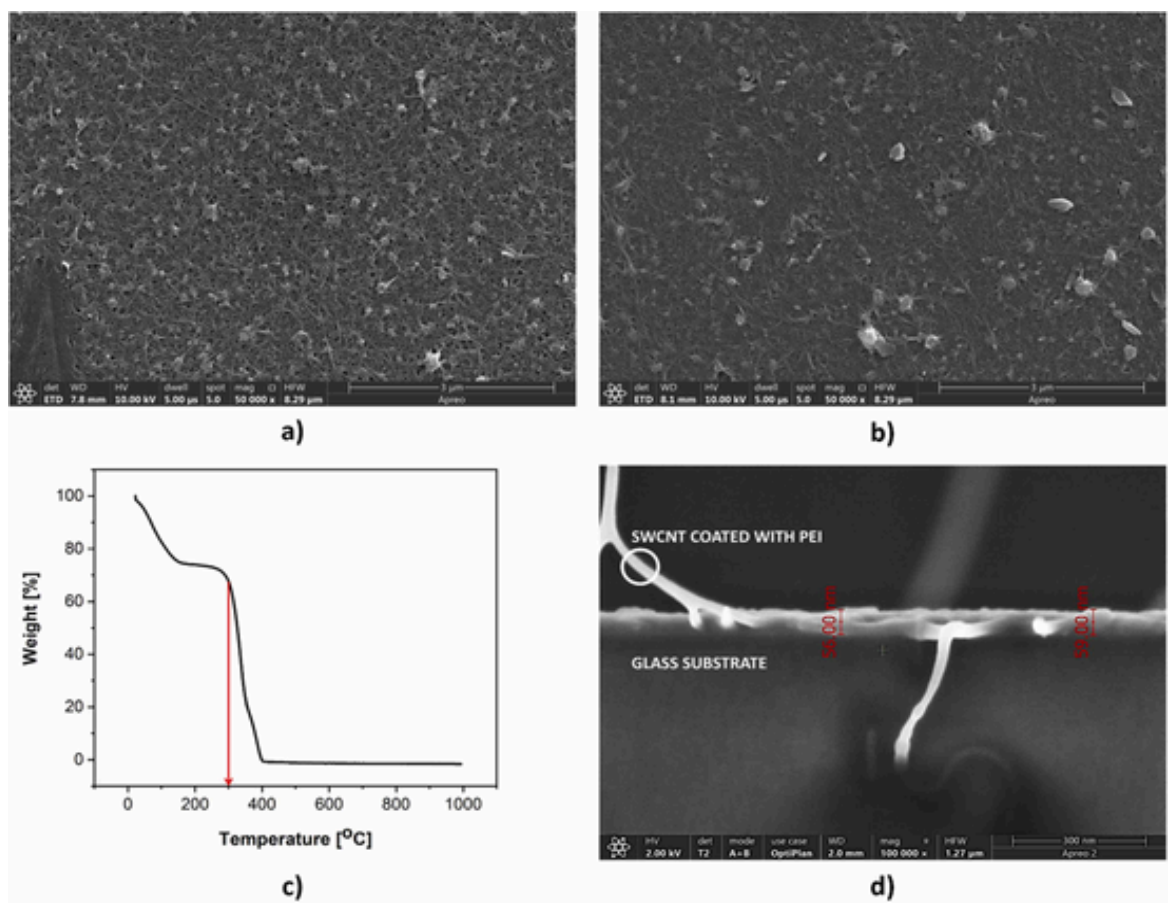

**Fig. 2.** a) HR SEM image of surface of the thermally untreated 10 bilayers PEI-SWCNT film, b) HR SEM image of surface of the thermally treated PEI-SWCNT (10) 300 film, c) TG/DSC analysis of PEI polymer and d) cross section of the thermally treated PEI-SWCNT film with 8 bilayers.

In Fig. 2d, the cross section of PEI-SWCNT (8) 300 is presented. A SEM image shows that the film is continuous, the adhesion between the substrate and the film is very good. The approximate thickness of 8 bilayers PEI-SWCNT film is about 60 nm which implies that the thickness of the 1 bilayer is roughly 7 nm. From the micrograph shown in Fig. 2d we measured the diameter of the isolated nanotube. An average diameter is around 20 nm, which is about 4–5 times larger than the value specified by the manufacturer. It is probably because each nanotube is coated with a PEI polymer, which affects the increase in diameter. The coating around each nanotube prevents direct contact between SWCNTs, which is expected to affect the electrical properties of the thin films. Exactly for that reason, we tried to improve electrode performance by thermal treatment at 300 °C. The following section first addresses the effect of such an approach on equally important optical properties.

### 3.2. Optical characterization

It is known that SWCNTs in the form of the thin film can be used as transparent electrodes, not only as a replacement for ITO in the visible range but also as transparent electrodes in the mid-IR range [1,3,29]. For this reason, transparency in a wide range of the electromagnetic spectrum is a very important property to be investigated.

#### 3.2.1. UV/VIS spectrophotometry

Fig. 3a and 3b show the dependence of transparency on wavelength before and after thermal treatment. The transparency of prepared thin films was measured in the spectral range from 400 to 900 nm. It can be noticed that the curves increase slightly towards the higher values of the wavelength and that the transparency tends to decrease with the number of layers. In Table 2 are values of transparency at 600 nm before and after thermal treatment, respectively. The results presented show higher transparency values after thermal treatment of approximately 7 %, which could be due to lower porosity and reduction in film thickness.

#### 3.2.2. FT-IR spectrophotometry

The FT-IR measurements in the mid-IR spectrum are also important for the application of the films in devices such as IR detectors. Fig. 3a shows the transparency of the sample series before thermal treatment. The dependence of transparency on wavenumber is non-linear and increases significantly after 3000 nm. The observed peaks (Fig. 3a) at 2922 cm$^{-1}$ and 2851 cm$^{-1}$ originated from PEI polymer [30–32]. On the other hand, Fig. 3b shows the FT-IR spectra after thermal treatment and PEI peaks cannot be seen. This is to be expected as the polymer slowly loses viscosity at 300 °C and the structure of the PEI partially begins to degrade. Table 3 shows the values of transparency at 3 μm, and just like the UV VIS results, the transparency values of the treated samples are slightly higher than those of the untreated samples. Based on the results it could be concluded that the PEI absorbs some fraction of the light, which affects the somewhat lower values of transparency.

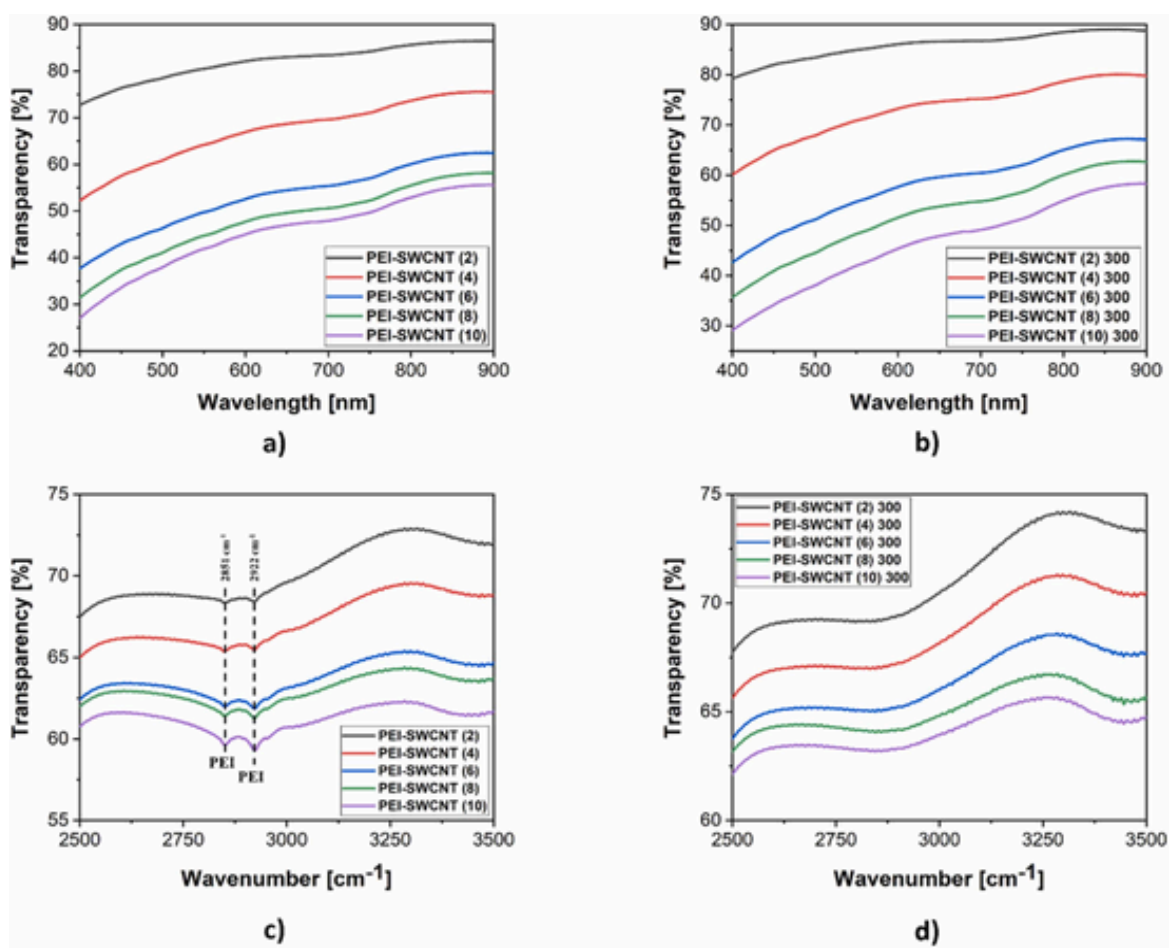

**Fig. 3.** a) UV/VIS spectra of PEI-SWCNT films composed of 2, 4, 6, 8 and 10 bilayers before thermal treatment and b) UV/VIS spectra of PEI-SWCNT films composed of 2, 4, 6, 8 and 10 bilayers after thermal treatment at the 300 °C for 1 h, c) FT-IR spectra of PEI-SWCNT films up to 10 bilayers before thermal treatment and d) FT-IR spectra of PEI-SWCNT films up to 10 bilayers after thermal treatment at the 300 °C for 1 h.

**Table 2**
Values of transparency at 600 nm of thermally untreated and treated samples.

| Sample label | Transparency at 600 nm [%] | | Sample label |
|---|---|---|---|
| | Thermally **untreated** samples | Thermally **treated** samples | |
| PEI-SWCNT (2) | 82 | 86 | PEI-SWCNT (2) 300 |
| PEI-SWCNT (4) | 67 | 73 | PEI-SWCNT (4) 300 |
| PEI-SWCNT (6) | 52.6 | 57.5 | PEI-SWCNT (6) 300 |
| PEI-SWCNT (8) | 48 | 51.5 | PEI-SWCNT (8) 300 |
| PEI-SWCNT (10) | 45 | 45 | PEI-SWCNT (10) 300 |

**Table 3**
Values of transparency at 3 μm of thermally untreated and treated samples.

| Sample label | Transparency at 3 μm [%] | | Sample label |
|---|---|---|---|
| | Thermally **untreated** samples | Thermally **treated** samples | |
| PEI-SWCNT (2) | 69.5 | 70.5 | PEI-SWCNT (2) 300 |
| PEI-SWCNT (4) | 66 | 68 | PEI-SWCNT (4) 300 |
| PEI-SWCNT (6) | 63 | 66 | PEI-SWCNT (6) 300 |
| PEI-SWCNT (8) | 62 | 65 | PEI-SWCNT (8) 300 |
| PEI-SWCNT (10) | 61 | 64 | PEI-SWCNT (10) 300 |

### 3.3. Electrical characterization

#### 3.3.1. Four-point probe measurements

The sheet resistance of PEI-SWCNT thin films composed of 2, 4, 6, 8 and 10 bilayers were measured with the four-point probe setup. The samples were measured at least three times at different locations along the surface of the sample and the average values of sheet resistivity for thermally non-treated and treated samples are given in Table 4. The sheet resistance of the non-treated samples was notably lower than the previously reported results with the 80 % purity nanotubes [22]. The lower sheet resistance observed in this study can be attributed to the use of higher-purity SWCNTs and possibly a more uniform distribution of nanotubes. This results in a network structure with increased interconnectivity among the nanotubes which affects the electrical properties. Also, the results show that the film resistance is even lower after the thermal treatment.

**Table 4**
Sheet resistance of PEI-SWCNT films composed of 2, 4, 6, 8 and 10 bilayers before and after thermal treatment.

| Sample label | Sheet resistance [kΩ/sq] | | Sample label |
|---|---|---|---|
| | Thermally **untreated** samples | Thermally **treated** samples | |
| PEI-SWCNT (2) | 15 | 4 | PEI-SWCNT (2) 300 |
| PEI-SWCNT (4) | 7.7 | 1.6 | PEI-SWCNT (4) 300 |
| PEI-SWCNT (6) | 8.4 | 1 | PEI-SWCNT (6) 300 |
| PEI-SWCNT (8) | 8.1 | 0.9 | PEI-SWCNT (8) 300 |
| PEI-SWCNT (10) | 7 | 0.8 | PEI-SWCNT (10) 300 |

In both cases (before and after thermal treatment), the transition from 2 to 4 bilayers had reduced sheet resistance strongly (Table 4). For sample series without thermal treatment resistance values are not consistent, probably due to random SWCNT ordering, reaching the conductivity saturation which is limited by the PEI coating. In the frame of this research, first was tried to use lower concentration of PEI to improve conductivity, but such an approach was not effective. A similar trend is noticed in the case of a thermally treated system, but with lower absolute values of sheet resistance. A continuous decrease of sheet resistance with number of deposited bilayers is observed, with a rather mild decrease of sheet resistance from 6 to 10 bilayers (Table 4, from 1 kΩ/sq to 0.8 kΩ/sq for thermally treated samples). Some relations between structure and electrical properties can be drawn, pointing out the effect of electrode exposure to 300 °C for 1 h. Namely, with the increase in the number of SWCNT layers, we obtained better electrical properties by increasing the number of contact points between nanotubes. Reaching out to assumptions posed in section *Structural characterization*, thermal treatment had a positive effect on electrode performance. Temperature indeed seemed to improve the contact between the nanotubes by densifying the system. It cannot be claimed that SWCNTs are in direct contact after exposure to 300 °C, but also not excluding such cases in some points of the PEI-SWCNT composite. It is more probable that the thickness of PEI medium between SWCNT is reduced. Bearing in mind that such a composite system is a very complex one from the point of understanding the conductivity mechanisms of the system, we have tried to study and solve this challenge by employing computational studies. Reported results provide a clear picture of the trade-off between optical and electrical properties. The optical characterization, as outlined in Table 2 and Table 3, reveals that higher transparency is achievable with a smaller number of bilayers, while the electrical characterization (Table 4) illustrates increased conductivity associated with an increased number of bilayers. Despite the superior transparency exhibited by the film with a 2 bilayer structure and the lowest sheet resistance observed in the film with 10 bilayer structure, our findings suggest that the 4 bilayer film represents optimal electrical and optical performances for the application a conductive electrode transparent in visible and IR spectrum.

### 3.4. Current-voltage properties

#### 3.4.1. Transport properties of SWCNT and SWCNT networks

Computational studies were done to investigate and understand the interplay between the added polymer layer and network of SWCNT nanotubes on the electrical properties of the film. Energy bandgap of semiconducting SWCNT can be estimated according to the following equation:

$$E_g = \frac{2\gamma_0 a_{cc}}{d_{SWCNT}}, \qquad (1)$$

where, $\gamma_0$ stands for tight binding energy between C atoms within the tube, $a_{cc}$ – carbon-carbon spacing, $d_{SWCNT}$ corresponds to SWCNT diameter. Considering that the average diameter of SWCNT within the samples is $d_{SWCNT} \approx 1.4 \pm 0.1 nm$, the value of tight binding energy $\gamma_0 = 3 eV$ and $a_{cc} = 0.142 nm$, energy bandgap can be estimated in the range from 0.57 eV – 0.65 eV in consisted with expected values for semiconductor nanotubes [33].

According to the specification given by the manufacturer existing bundles within a nanotube network (NTN) have average diameter $D \approx 4 - 5 nm$ and the average length in the range from $l \approx 0.5 - 5 \mu m$. Considering the average length, a single SWCNT is out of ballistic transport regime and the SWCNT resistance is linearly dependent on the length of conductive path mainly governed by the diffusive transport and scattering induced by phonons [34,35].

SWCNT-PEI junction is confirmed to have a highly positive charged surface [36] which leads to the formation of a condensed layer determined by the Bjerrum length of PEI (for the dielectric constant of PEI $\varepsilon_r = 4$ Bjerrum length equals to $l_B = 14 nm$). Two likely charged rod-like SWCNT-PEI structures can exhibit strong attraction (long-range Coulomb forces) due to mutual polarization as well as randomization of ions within the condensation layer and conversion to a less ordered state [36]. This effect can also lead to issues with the reproducibility of experimental results [36].

#### 3.4.2. Results of computational studies and discussion

Linear I-V characteristics, shown in Fig. 4a is identified as Ohmic behavior indicating quasi-metallic conductivity within the layers [32]. This behavior is in agreement with the fact that PEI layer leads to n-type conductivity within CNTs, also proved by Hall measurement results (Supporting Information, Table S1 and Table S2). Moreover, linearity in current-voltage characteristics of CNTs is also assigned to uniformity of transmission spectra of CNTs which determines current values according to Landauer-Büttiker formula [37].

The decrease of overall resistance with number of layers, from 22.7 kΩ for 10 layer film to 2.27 kΩ (Supporting Information, Figures S1 - S5) for two-layer film originates from higher SWCNT density and reduced phonon influence [35].

We assume that overall film resistance composed of a SWCNTs network is strongly influenced by the resistance of junctions [38], formed between SWCNT and PEI layer as well as intertube and interbundle junctions. Other numerous effects that also contribute to charge transport include: scattering on defects, transport along the tube, energy profile of dielectric tube junction and different sizes of tubes, etc., which leads to absence of an analytical, or numerical model which takes into account all of the possible interplays [35]. Additionally, in single wall NTN, hopping between junctions of different nanotubes is likely to occur, as well as tunneling through different junctions [35] in most cases thermally assisted tunneling [39].

Current-voltage characteristics for samples exposed to thermal treatment up to 300°C (PEI-SWCNT (10) 300) and samples not thermally treated (PEI-SWCNT (10)) is shown in Fig. 4b. The resulting graphs indicate that overall sample resistance diminishes significantly from 145.12 kΩ to 43.65 kΩ for samples treated at 300 °C. As already mentioned, temperature treatment applied during fabrication procedure decreases thickness of PEI layer between the tubes, leading to the higher density of the NTN. Higher SWCNT density reduces phonon scattering and increases overall currents (Fig. 4b) [35]. Moreover, PEI layer can be represented as a potential barrier between two SWCNT, which reduction results in larger values of tunneling probability. Namely, it is

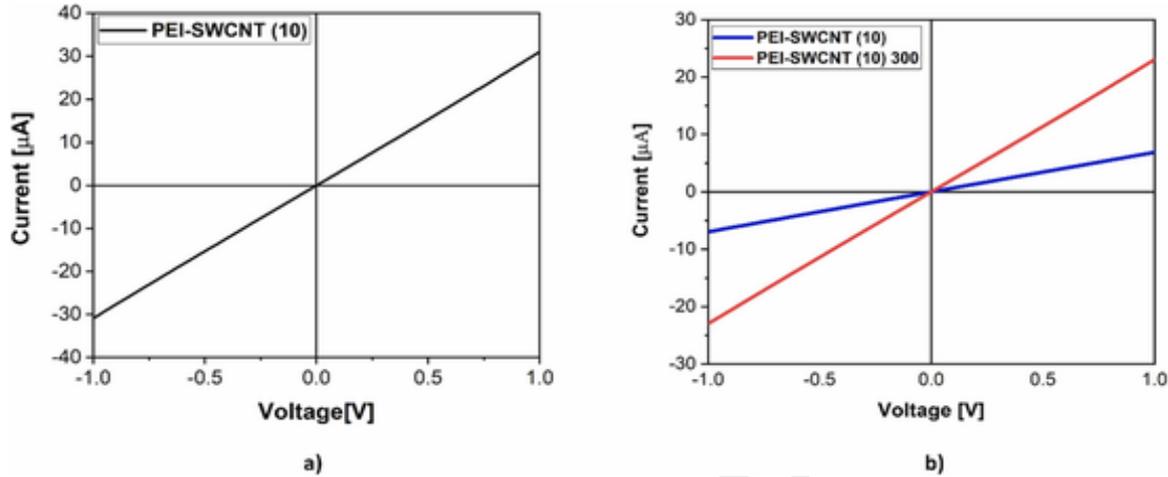

**Fig. 4.** a) I-V characteristic of a PEI-SWNCT (10) film and b) linear I-V characteristics of a 10 bilayers films before (PEI-SWNCT (10)) and after (PEI-SWNCT (10) 300) thermal treatment.

well known that tunneling probability $T$ exponentially decays with barrier width $L$:

$$T \sim e^{-2\alpha L} \qquad (2)$$

where parameter α depends on barrier high $V_0$ and energy of incident electrons E, and effective mass of electrons $m^*$, $\alpha = \sqrt{\frac{2m^*(V_0-E)}{\hbar^2}}$. Assuming that PEI-SWCNT energy barrier profile remains unchanged before and after thermal treatment, we can expect exponential decay of tunneling probability with thicker PEI layers.

Experimental results shown in Fig. 5 show decrease of conductivity when samples are heated above room temperature, up to 200 °C. This behavior is assigned to electron-phonon scattering transport mechanisms, with higher phonon scattering rate on elevated temperature [35]. The referred effect leads to reduction of electron mobility within the single tube of diameter [$d$] exposed to temperature [$T$], which can be described with empirical relation:

$$\mu(T,d) = \mu_0 \frac{300K}{T} \left(\frac{d}{1nm}\right)^{\beta} \text{[35]}, \qquad (3)$$

where is $\beta = 2.24$ and $\mu_0 = 12000\ cm^2 V^{-1} s^{-1}$. Reduction of charge carrier mobility directly influences reduction of overall sample conductivity. It is worth noting that for NTN scattering at inter-nanotube junctions should also be considered. Such behavior also indicates that tunnelling mechanisms between nanotubes, which are enhanced with the temperature, do not play a major role in material conductivity since temperature increase lowered conductivity.

## 4. Conclusion

This research was focused on optimizing optical and electrical properties of PEI-SWCNT thin films through deeper understanding of the polymer effect and the presence of PEI-SWCNT junctions. The Raman spectrometry confirmed the existence of stable, undisrupted high-quality SWCNT structure, while the HRSEM showed relatively uniform distribution of SWCNT with significant surface porosity before the thermal treatment at 300 °C. After treatment, which is supported by TG/DSC analysis, we can conclude that the structure is denser due to the polymer chains flow leading to reduction of the porosity. The HRSEM cross section showed that the thickness of the 8 bilayer film is around 60 nm, and that the PEI polymer coats the nanotubes, preventing direct contact between them. Exactly for this reason the samples were thermally treated to achieve a closer and more direct contact between the nanotubes in a denser structure. Optical characterization in the UV/VIS spectra showed higher transparency in comparison to our previous work [22], while there were no significant differences for the mid-IR spectra. Namely, the transparency in the UV/visible spectra was 45–82 % and 45–86 % for untreated and thermally treated samples, respectively. In the mid-IR spectra, the transparency values were between 61 % and 69.5 % and 64–70.5 % for untreated and thermally treated samples, respectively. The sheet resistance for untreated samples was 7–15 kΩ/sq and for thermally treated samples was 0.8–4 kΩ/sq, reducing with more layers deposited. Lower sheet resistance after 300 °C confirmed the effect of reduced distance between the tubes which is also in agreement with the HRSEM results. Computational results showed that the main influence results primarily from the resistance of junctions, formed between SWCNT and PEI layer, as well as intertube and interbundle junctions and that the temperature-enhanced tunnelling mechanisms between nanotubes do not play a major role in material conductivity. As a general conclusion, we can estimate that the optimal film design for the electrode application is PEI-SWCNT (4) 300, because further increase after 4 deposited bilayers increases electrical conductivity only to a minor extent and reduces the transparency in mid-IR range which is undesirable.

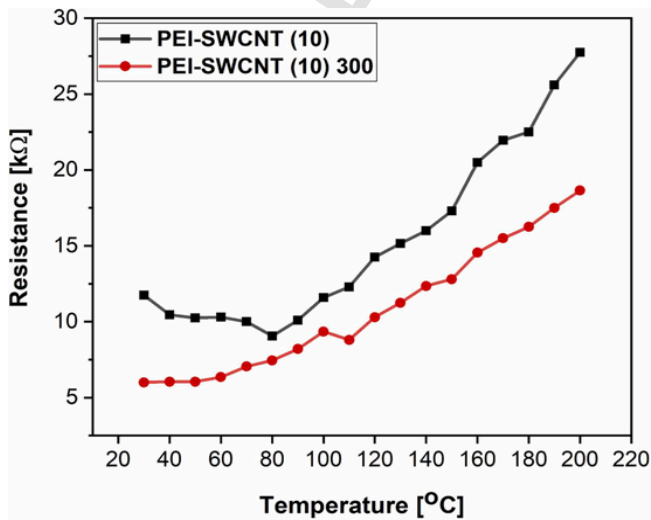

**Fig. 5.** Trend of resistance with the temperature for PEI-SWCNT (10) and PEI-SWCNT (10) 300.

**Uncited references**

[14,15,17–21]

**CRediT authorship contribution statement**

**Branimir Bajac:** Writing – review & editing, Writing – original draft, Supervision, Methodology, Conceptualization. **Nataša Samardžić:** Writing – review & editing, Writing – original draft, Validation, Investigation, Conceptualization. **Jovana Stanojev:** Writing – review & editing, Supervision, Methodology, Conceptualization. **Sara Joksović:** Writing – review & editing, Writing – original draft, Investigation.

**Declaration of Competing Interest**

The authors declare that they have no known competing financial interests or personal relationships that could have appeared to influence the work reported in this paper

**Acknowledgments**

The authors want to express their gratitude to the prof. dr Akos Kukovecz and prof. dr Imre Szenti from the Department of Applied and Environmental Chemistry, University of Szeged for performing HR SEM of the samples. This work is supported through the ANTARES project that has received funding from the European Union's Horizon 2020 research and innovation programme under grant agreement SGA-CSA. No. 739570 under FPA No. 664387. https://doi.org/10.3030/739570 and Ministry of Science, Technological Development and Innovation of Republic of Serbia 451-03-66/2024-03/200358.

**Data availability**

Data will be made available on request.

**Appendix A. Supporting information**

Supplementary data associated with this article can be found in the online version at doi:10.1016/j.synthmet.2024.117717.